\documentclass{article}

\PassOptionsToPackage{numbers, compress}{natbib}

\usepackage[preprint]{neurips_2026}

\usepackage[utf8]{inputenc} 
\usepackage[T1]{fontenc}    
\usepackage{hyperref}       
\usepackage{url}            
\usepackage{booktabs}       
\usepackage{amsfonts}       
\usepackage{nicefrac}       
\usepackage{microtype}      
\usepackage{xcolor}         
\usepackage{amsmath}
\usepackage{graphicx}
\usepackage{wrapfig}
\usepackage{subcaption}
\usepackage{enumitem}
\usepackage{comment}
\usepackage{xspace}
\usepackage{titlesec}
\titlespacing*{\paragraph}{0pt}{0.6ex plus 0.2ex minus 0.1ex}{0.6em}

\definecolor{dong}{RGB}{0, 0, 200}
\definecolor{han}{RGB}{0, 200, 0}
\definecolor{checked}{RGB}{0, 0, 0}

\newcommand{\name}{ChunkFlow\xspace}

\title{ChunkFlow: Communication-Aware Chunked Prefetching for Layerwise Offloading in Distributed Diffusion Transformer Inference}

\author{%
  Han Meng \\
  University of California, Merced \\
  \And
  Danny Willow Liu \\
  University of Chicago \\
  \And
  Dong Li \\
  University of California, Merced \\
  Yotta Labs \\
}

\begin{document}

\maketitle

\begin{abstract}
Layerwise offloading reduces the GPU memory footprint of large diffusion
transformer (DiT) inference by prefetching upcoming layers from host memory,
but its effectiveness hinges on hiding prefetch latency behind per-layer
computation. This assumption breaks down when the per-GPU compute workload is small. Moreover, on PCIe-only nodes, prefetch and inter-GPU collective communications such as all-reduce and all-to-all contend on the shared PCIe path, exposing prefetch latency even when compute would otherwise hide it. We revisit layerwise offloading as a
co-scheduling problem between prefetch and communication, guided by a
first-order analytical model that predicts when prefetch can be hidden by computation. Building on this model, we design \name, a
communication-aware, chunk-granular offloading runtime that adaptively yields
to collective communication and smoothly trades GPU memory for prefetch volume. On
three representative diffusion transformers running on two H100 GPUs over
PCIe with Ulysses sequence parallelism, \name delivers up to
$1.28\times$ step-time speedup over SGLang's
existing layerwise offloading, reduces peak GPU memory by up to $49\%$ over
the no-offload baseline at near-identical step time once the workload is
large enough, and exposes a tunable memory--latency tradeoff that recovers near-zero step-time overhead in the small-workload regime. 
\end{abstract}

\section{Introduction}
\label{sec:intro}

Diffusion models~\cite{ddpm2020,scoresde2021} have become one of the dominant
paradigms for generative modeling, powering state-of-the-art systems in
image~\cite{ldm2022,sdxl2024} and video~\cite{sora2024,hunyuanvideo2024,wan2025}
generation. Modern systems increasingly adopt Transformer-based
denoisers~\cite{attention2017,vit2021}, commonly referred to as Diffusion
Transformers (DiTs)~\cite{dit2023,pixart2024,mmdit2024}, which scale to
tens of billions of parameters and unify architectures across modalities. A key bottleneck in
serving large DiTs is GPU memory capacity.
\textcolor{checked}{Unlike LLMs, where sequence lengths are typically on the order of $10^3$–$10^4$ tokens, DiTs operate on significantly longer sequences due to spatial and temporal tokenization, often reaching $10^5$ to $10^6$ tokens for high-resolution images and videos. For example, generating a single $1280{\times}720{\times}129$-frame video with the 13B-parameter HunyuanVideo~\cite{hunyuanvideo2024} requires distributing the model across $8$ H100 GPUs via sequence parallelism using the official xDiT~\cite{xdit2024} configuration. This scaling is primarily driven by the memory footprint of large model weight and long-sequence attention—dominated by activations and KV buffers—rather than intrinsic compute demand.}

A standard remedy to the above problem is \emph{layerwise offloading}, which stores weights in
host memory and asynchronously prefetches them \textcolor{checked}{to the GPU memory one layer ahead}. 
By keeping only two layers on the GPU memory, this solution substantially reduces peak
memory, and hence is adopted in production-quality inference systems such as
SGLang~\cite{sglang2024} and vLLM~\cite{kwon2023efficient}.

\textcolor{checked}{The efficiency of layerwise offloading depends on whether the per-layer compute time is long enough to hide the host-to-device (H2D) prefetch time. Whether this overlap is achievable is determined by per-GPU compute throughput and host-to-device bandwidth. When the overlap is achievable, offloading reduces memory without inflating latency; otherwise, the unfinished prefetch is exposed to the critical path and slows down inference.}

\begin{figure}[!t]
\centering
\begin{minipage}[b]{0.50\linewidth}
    \centering
    \includegraphics[width=\linewidth]{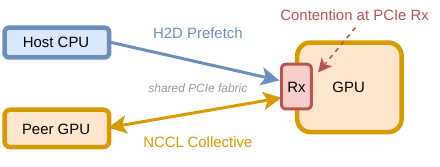}
    \subcaption{Data-path sharing.}
    \label{fig:pcie-contention}
\end{minipage}\hfill
\begin{minipage}[b]{0.50\linewidth}
    \centering
    \includegraphics[width=\linewidth]{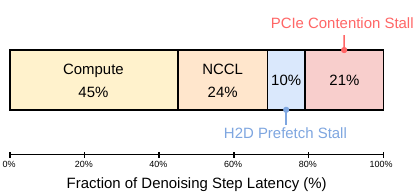}
    \subcaption{Latency breakdown.}
    \label{fig:latency-breakdown}
\end{minipage}
\caption{PCIe contention on PCIe-based nodes during distributed DiT
inference. (a) H2D prefetch and inter-GPU collectives converge at each
GPU's PCIe receive port (Rx) and contend for bandwidth. (b) Denoising
step latency breakdown: the contention stall ($21\%$) accounts for a large portion.}
\vspace{-20pt}
\label{fig:contention-overview}
\end{figure}

\textcolor{checked}{In commodity multi-GPU deployments---e.g., 8$\times$L40 or 8$\times$A6000 PCIe nodes widely used in cloud and on-premise environments~\cite{jiang2025heterogpu,hgca2025,gpucomm2024}---intra-node GPU-to-GPU traffic traverses PCIe rather than a dedicated high-bandwidth fabric (NVLink). On such PCIe-only nodes, distributed inference introduces an additional issue beyond the baseline overlap analysis above: H2D prefetch and inter-GPU collective communications---such as all-reduce, all-gather, and all-to-all, commonly employed in distributed inference for weight and activation re-sharding (e.g., Ulysses sequence parallelism~\cite{ulysses2023})---traverse the same PCIe fabric and converge at each GPU's PCIe receive port (Rx), as illustrated in Figure~\ref{fig:pcie-contention}. This \textit{data-path sharing} can significantly slow down the prefetching.} 
\textcolor{checked}{For example, on two H100 GPUs over PCIe with Ulysses sequence parallelism, enabling existing layerwise offloading at the official default resolutions inflates denoising step time by up to $2.0\times$, $1.9\times$, and $1.8\times$ on WanVideo (5B)~\cite{wan2025}, HunyuanVideo (13B)~\cite{hunyuanvideo2024}, and Flux (12B)~\cite{flux2024}, respectively.}
\textcolor{checked}{The data-path sharing problem is the major bottleneck that prevents the combination of the offloading and parallel strategy to maximize GPU memory saving while having short inference latency.}

In this work, we build a first-order analytical model to decide when layer prefetch can be hidden behind per-block computation. The model defines a single quantity---the
\emph{critical compute workload} $F^\star$---that characterizes the regime boundary: when the block-level FLOPs $F_{\text{block}}$ exceed $F^\star$, full-layer prefetching is expected to incur little overhead; otherwise,
part of the prefetch is exposed. Guided by this model, we design \textcolor{checked}{\name, a chunk-granular offloading runtime} with two mechanisms: (1) communication-aware chunked prefetching, which splits each layer's parameters into fixed-size chunks and enables pauseable/resumable transfers of weights. Based on this mechanism, the collective communications can be triggered between chunk boundaries, instead of being
blocked behind a full-layer transfer of parameters. The contention slowdown on PCIe-based nodes is effectively reduced. (2) Chunk-granular partial parameter residency, which keeps a tunable fraction of chunks resident in the GPU memory
and prefetches the rest, providing a \textcolor{checked}{tunable, fine-grained} memory--latency trade-off for workloads below $F^\star$.


\textcolor{checked}{On} representative DiTs spanning text-to-image and text-to-video generation, \textcolor{checked}{\name} consistently reduces denoising step time over state-of-the-art layerwise offloading for DiT \textcolor{checked}{by up to $1.28\times$}, matches the no-offload baseline (\textcolor{checked}{assuming the GPU memory is sufficient}) once the workload is large enough while retaining \textcolor{checked}{up to $49\%$} peak-memory savings. 


\paragraph{Contributions.} \textcolor{checked}{(1) We develop a first-order analytical model that characterizes when layer prefetch can be fully hidden behind per-layer compute, applicable to layerwise offloading in general. (2) We identify a previously overlooked failure mode during layerwise offloading on PCIe-based distributed execution---contention between layer prefetch and collective communication on the shared PCIe path---and address this problem with a communication-aware chunked prefetching mechanism that yields PCIe to collectives at chunk boundaries. (3) For workloads whose per-layer compute is intrinsically too small to hide prefetch, we introduce a chunk-granular partial parameter residency mechanism that provides a tunable, fine-grained memory--latency trade-off.}

\section{Background \& Motivation}
\label{sec:background}

\subsection{DiT Inference and Distributed Execution}
\label{subsec:background-dit}

\textcolor{checked}{DiTs~\cite{dit2023} couple an iterative sampler with a Transformer-based denoiser invoked once per denoising step on a sequence of $S$ tokens obtained after VAE downsampling and patchification.}

To scale beyond a single GPU, distributed DiT inference commonly employs
tensor parallelism (TP) and sequence (or context) parallelism (SP).
TP
partitions weight matrices and attention heads across devices, and reconciles
partial results within a block through collective communications such as all-reduce or
all-gather, depending on the sharding pattern. SP partitions tokens across
devices and exchanges them through all-to-all communication inside attention;
\textcolor{checked}{in particular, Ulysses SP~\cite{ulysses2023} has been widely adopted as the SP backbone in distributed DiT inference engines~\cite{pipefusion2024,xdit2024} and is the SP setup we use throughout this paper.}
These collective communications are interleaved with computation within each block and can
lie directly on the critical path of execution.

\subsection{Layerwise Offloading}
\label{subsec:background-offload}

Layerwise offloading reduces GPU memory footprint by keeping only a small working set
of layers resident on the GPU, while the remaining weights are staged in host
memory and transferred back on demand. Offloading is performed at layer
granularity: during initialization all layers are placed in pinned host memory,
and each denoising step then iterates over Transformer blocks.
\textcolor{checked}{The processing of each layer $l$ proceeds as follows.}
\vspace{-5pt}
\begin{itemize}[itemsep=0pt, topsep=2pt,leftmargin=*]
    \item \textcolor{checked}{The compute stream waits for the asynchronous prefetch of the layer $l$ to complete, ensuring its parameters are resident on the GPU.}
    \item \textcolor{checked}{The compute stream then executes the forward pass of the layer $l$.}
    \item \textcolor{checked}{During this execution, a dedicated CUDA copy stream asynchronously prefetches the next layer ($l{+}1$).}
    \item \textcolor{checked}{After the layer $l$ completes, the runtime releases the parameters in $l$ to reclaim GPU memory.}
\end{itemize}
\vspace{-5pt}

This reduces the resident working set to only two layers (one active, and one to prefetch). The standard implementation hinges on the assumption that
the prefetch of layer $l{+}1$ finishes before layer $l$'s compute does;
otherwise, the computation stream must explicitly wait for the prefetch
stream at step (1), introducing stalls on the critical path.

\subsection{PCIe Contention under Distributed Offloading}
\label{subsec:background-communication-path}

The interaction between offloading and communication depends strongly on the
interconnect. On NVLink-equipped systems, inter-GPU collectives use a dedicated
high-bandwidth fabric while H2D transfers use the PCIe I/O path, so the two do
not directly interfere. On PCIe-only systems\textcolor{checked}{---commonplace in commodity multi-GPU servers (e.g., 8$\times$L40 PCIe nodes) that underpin budget-conscious DiT deployments---}%
 both inter-GPU collectives and
H2D prefetch traverse the same PCIe fabric and converge at each GPU's PCIe
receive port, creating the potential for cross-traffic contention.

\textbf{Performance characterization.} We evaluate this effect using a 5B-parameter WanVideo diffusion transformer on a
single node with two H100 GPUs connected via \textcolor{checked}{PCIe Gen5 $\times$16 (peak $64$\,GB/s per-GPU Rx shared between CPU-to-GPU H2D transfers and inter-GPU collective traffic)},
generating 81 frames with Ulysses
sequence parallelism~\cite{ulysses2023}. 
Enabling layerwise offloading increases the average
denoising step latency from $1.307$\,s to $1.888$\,s---a $1.44\times$
slowdown. The intrinsic cost of the all-to-all communication in
sequence-parallel attention is an order of magnitude smaller than this gap,
indicating that the slowdown cannot be explained by raw PCIe saturation alone.
\textcolor{checked}{Figure~\ref{fig:latency-breakdown} breaks down the inflated step latency:
the NCCL all-to-all collective itself accounts for $24\%$, while the PCIe-contention
stall---the all-to-all delayed behind in-flight prefetch traffic at the GPU's Rx---accounts
for $21\%$, an overhead comparable to the collective's own time and confirming that the contention, rather than raw bandwidth or communication cost, dominates
the gap.}

Profiling traces reveal the underlying mechanism. In the default
implementation, the prefetch stream for the next layer is issued near the
beginning of the current layer and therefore precedes the in-block
all-to-all. By the time the compute stream reaches the communication phase,
the GPU Rx path is already occupied by an in-flight H2D prefetch. Because
PCIe provides no prioritization between DMA transfers and collective traffic,
the all-to-all is effectively delayed until the prefetch transfer completes,
and the prefetch latency that was supposed to be hidden by computation is
instead exposed to the critical path.

\section{Design}
\label{sec:design}

\textcolor{checked}{This section addresses three layered questions. Section~\ref{subsec:design-overlap-model} asks \emph{when full-layer prefetch can in principle be hidden behind per-block computation}, deriving a first-order overlap model with a critical compute workload $F^\star$. The two subsequent mechanisms then tackle two distinct failure modes identified by the model. Section~\ref{subsec:design-chunked-prefetch} (PCIe-specific) recovers overlap when prefetch is theoretically hidable ($F_{\text{block}} \ge F^\star$) but PCIe contention with collectives breaks it. Section~\ref{subsec:design-partial-residency} (interconnect-agnostic) trades GPU memory for prefetch volume when overlap is intrinsically infeasible ($F_{\text{block}} < F^\star$).}

\subsection{Analytical Model for Prefetch--Compute Overlap}
\label{subsec:design-overlap-model}

We build an analytical model for two dominant terms in one block execution window: the computation time of the current block and the host-to-device prefetch time of the next block. \textcolor{checked}{This model captures the first-order prefetch--compute overlap behavior of layerwise offloading: it depends only on per-GPU compute throughput and H2D bandwidth, and therefore applies regardless of the interconnect (PCIe or NVLink) and to both single-GPU and distributed settings.}

\paragraph{Compute model.}
We model the per-block compute cost of one Transformer block on a
single GPU. Our evaluation covers two block designs: the standard
\emph{DiT} block (e.g., in WanVideo), which serially interleaves
self-attention, cross-attention to the text context, and an MLP; and
the \emph{MM-DiT} block~\cite{mmdit2024} (e.g., in Flux,
HunyuanVideo), which jointly processes image and text streams in a
single attention call. We focus on the high-FLOP operators (attention
projections, attention, and the MLP); lower-order components such as
normalization, softmax, and elementwise operations are absorbed into
an empirical calibration factor introduced below. The explicit
per-term FLOP derivations for both block types are deferred to
Appendix~\ref{app:block-compute}.

Let $B$ denote the batch size, $S$ the global sequence length after
patchification, $d$ the hidden dimension, $f$ the FFN expansion
dimension, and $L_{\text{ctx}}$ the text context length. We write the
resulting per-block FLOP count as $F_{\text{block}}(B, S)$. Under
fixed $(d, f, L_{\text{ctx}})$, $F_{\text{block}}$ scales linearly in
$B$ and is quadratic in $S$ only through self-attention while all
other terms are linear in $S$.

\textcolor{checked}{Given the peak BF16 throughput $P_{\text{peak}}$ of the GPU, we model the block compute time as
\begin{equation}
T_{\text{comp}}(B, S) = \frac{F_{\text{block}}(B, S)}{\eta_{\text{comp}} P_{\text{peak}}},
\end{equation}
where $\eta_{\text{comp}} \in (0,1]$ is a compute calibration factor capturing the realized fraction of peak BF16 throughput on the target hardware (absorbing imperfect tensor-core utilization, memory-bound kernels, and non-GEMM operators); it is calibrated empirically per platform.}

\paragraph{Prefetch model.}
\textcolor{checked}{Let $B_{\text{pref}}$ denote the parameter bytes transferred before the next layer can execute (the entire layer for standard layerwise offloading) and $BW_{\text{h2d}}$ the host-to-device bandwidth. We model the prefetch time as
\begin{equation}
T_{\text{pref}} = \frac{B_{\text{pref}}}{\eta_{\text{pref}} BW_{\text{h2d}}},
\end{equation}
where $\eta_{\text{pref}} \in (0,1]$ is a prefetch calibration factor absorbing DMA startup cost, transfer granularity effects, and software runtime overheads; it is similarly calibrated empirically.}

\paragraph{Critical compute workload.}
The first-order condition for fully hiding prefetch behind computation is defined as follows. 
\begin{equation}
T_{\text{comp}} \ge T_{\text{pref}}.
\end{equation}
This inequality states that the computation window of the current block must be
at least as long as the service time of the next-layer prefetch.

Substituting the compute and prefetch models above, the overlap condition can
be written directly in terms of the block-level compute workload
$F_{\text{block}}$. 
\begin{equation}
F_{\text{block}}(B, S) \ge F^\star,
\qquad
F^\star
\;:=\;
\eta_{\text{comp}} P_{\text{peak}} \cdot T_{\text{pref}}
=
\frac{\eta_{\text{comp}} P_{\text{peak}} B_{\text{pref}}}
     {\eta_{\text{pref}} BW_{\text{h2d}}}.
\end{equation}
We call $F^\star$ the \emph{critical compute workload}: when
$F_{\text{block}} \ge F^\star$, the compute window is long enough to hide a
full prefetch; when $F_{\text{block}} < F^\star$, a part of the transfer is
exposed to the critical path. Equivalently, $I^\star := F^\star / B_{\text{pref}}$
is the turning point between the calibrated GPU compute roof and the
host-to-device bandwidth roof in a roofline plot, connecting our analysis
to the Hierarchical Roofline Model~\cite{moelightning2024}; see
Appendix~\ref{app:roofline} for the graphical view\textcolor{checked}{, and Appendix~\ref{app:calibration} for the numerical instantiation of $F^\star$ on our evaluation platform}.

Since $B_{\text{pref}}$ is approximately fixed for a given layer while $F_{\text{block}}$ grows monotonically in both $B$ and $S$, increasing either $B$ or $S$ pushes the workload across $F^\star$ and is generally favorable; conversely, partitioning computation across more GPUs (e.g., higher tensor or sequence parallelism degree) shrinks the per-GPU $F_{\text{block}}$ and can pull the local workload back below $F^\star$.

\subsection{Communication-Aware Chunked Prefetching}
\label{subsec:design-chunked-prefetch}

\begin{figure}[!t] \centering \includegraphics[width=\linewidth]{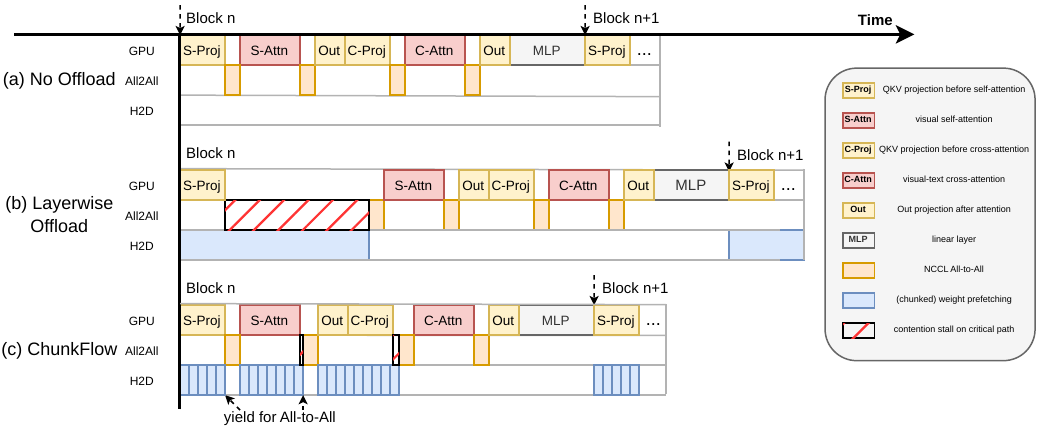} \caption{Block-level timing under (a) no offload, (b) whole-layer layerwise offload, and (c) our communication-aware chunked prefetch, illustrated on a DiT block with Ulysses sequence parallelism.} \label{fig:timeline} 
\vspace{-20pt} 
\end{figure}

In the default whole-layer prefetch design, the prefetch of layer $l{+}1$ is
issued near the beginning of layer $l$ on a dedicated copy stream.
However, the computation window before the first collective communication in a transformer block is typically short---for instance, in a self-attention
layer under Ulysses sequence parallelism the collective is needed \textcolor{checked}{right} after the QKV projection.
As a result, next-layer prefetch and collective communication frequently overlap in time and contend for the same PCIe path on PCIe-only nodes. Once a large H2D transfer is in flight, the later-arriving collective is
effectively delayed until the ongoing transfer finishes, so a part of the prefetch latency that is intended to be hidden by computation is instead
exposed to the critical path, as illustrated in Figure~\ref{fig:timeline}(b).

To address this issue \textcolor{checked}{in PCIe-based environments}, we design a communication-aware chunked prefetching
mechanism. The key idea is to make layer prefetch \emph{pauseable} and
\emph{resumable}, so that prefetch traffic can temporarily yield to the collective
communications. Rather than issuing the entire next layer as one monolithic
transfer, we partition the parameters of each layer into fixed-size chunks and
prefetch them asynchronously chunk by chunk on the copy stream.

\textcolor{checked}{Since PCIe provides no mechanism to prioritize collective traffic over in-flight prefetch (Section~\ref{subsec:background-communication-path}) and CUDA streams provide no built-in preemption either, we orchestrate pause/resume in software via a lightweight control flag checked at chunk boundaries. When the compute stream issues an inter-GPU collective, the runtime sets a per-layer pause flag and records a CUDA event signaling the collective's completion. Before launching each chunk's H2D copy, the prefetch worker checks the flag; if set, it enqueues a \texttt{cudaStreamWaitEvent} on the recorded event so that the copy stream stalls on the GPU until the collective finishes, after which subsequent chunks resume automatically. This design preserves correctness---no in-flight DMA is ever aborted---and lets the collective interleave with chunk prefetching instead of serialized behind a full-layer transfer, as shown in Figure~\ref{fig:timeline}(c). Because the copy stream can only yield at chunk boundaries, a small residual stall---bounded by the service time of one in-flight chunk---remains, but much shorter than the full-layer blocking of whole-layer prefetch.}

\textcolor{checked}{Chunking does not change the total I/O volume---it only turns a monolithic transfer into a sequence of shorter ones, leaving the chunk size $C$ as a tunable parameter. Concretely, the residual stall above equals one chunk's service time $T_{\text{chunk}} = C / (\eta_{\text{pref}} BW_{\text{h2d}})$, and $C$ governs a tradeoff between transfer efficiency and pause responsiveness: larger $C$ amortizes DMA startup and runtime overhead and improves H2D efficiency at the cost of a longer residual stall, while smaller $C$ enables finer-grained interruption but underutilizes PCIe bandwidth.}

\subsection{Reducing Prefetch Time via Partial Parameter Residency}
\label{subsec:design-partial-residency}

\textcolor{checked}{When $F_{\text{block}} < F^\star$, full offloading unavoidably exposes part of the prefetch latency on the critical path, creating an inherent tradeoff between GPU memory footprint and exposed prefetch volume.}

To navigate this tradeoff, we introduce \emph{partial parameter residency}\textcolor{checked}{, which, unlike the chunked prefetching mechanism in Section~\ref{subsec:design-chunked-prefetch}, is not specific to PCIe-based nodes and applies whenever the per-layer compute window is too short to hide a full prefetch}. Instead of offloading all parameter chunks of a layer, we keep a
subset of chunks resident in GPU memory and prefetch only the remaining ones.
\textcolor{checked}{This mechanism is a natural extension to the chunked layout introduced in Section~\ref{subsec:design-chunked-prefetch}:} since \name already partitions each layer into fixed-size chunks, the residency can be controlled at chunk granularity rather than at the coarser
module or whole-layer granularity. This enables finer control of the
memory--latency tradeoff.

\textcolor{checked}{This continuously trades prefetch volume for memory: at $0\%$ residency the mechanism reduces to full offloading, at $100\%$ residency to no offloading, and intermediate levels recover overlap when $F_{\text{block}} < F^\star$ at the cost of additional resident memory.}

\section{Evaluation}
\label{sec:eval}

\subsection{Experimental Setup}
\label{subsec:eval-setup}

We evaluate three open-source diffusion transformers spanning two
modalities: Wan2.2-TI2V-5B~\cite{wan2025} (5B, text-to-video), HunyuanVideo~\cite{hunyuanvideo2024} (13B, text-to-video), and FLUX.1-dev~\cite{flux2024} (12B, text-to-image). For each model we use the official default resolution, $10$ denoising steps, and guidance
configuration, and use a single fixed prompt across all runs.

Experiments run on a single node with two NVIDIA H100 GPUs over PCIe, using
Ulysses sequence parallelism~\cite{ulysses2023} with degree $2$ and
FlashAttention~\cite{flashattention2022,flashattention2_2023,flashattention3_2024}
as the attention backend. \name 
is built on SGLang~\cite{sglang2024} multimodal-generation runtime. We compare three
configurations: (1) No Offload (all weights resident, assuming \textcolor{checked}{GPU memory is sufficient, which represents the shortest inference latency)}, (2) layerwise offload (SGLang's whole-layer prefetching, \textcolor{checked}{named Layerwise in the later discussion}), and (3) \name.  

\textcolor{checked}{The overlap model and both the mechanisms generalize beyond this specific GPU count and parallelism style. Varying the GPU count simply re-parameterizes the per-GPU $F_{\text{block}}$ and $B_{\text{pref}}$ in the overlap model (Section~\ref{subsec:design-overlap-model}). Switching to a different parallelism style---e.g., tensor parallelism or other SP variants---only shifts where collectives are inserted within each block, and the collective type itself (all-reduce, all-gather, and all-to-all) does not change how the chunked prefetch worker yields: all interact identically through the pause/resume control flag discussed in Section~\ref{subsec:design-chunked-prefetch}.} \textcolor{checked}{The mechanisms are also architecturally portable beyond DiT, but the underlying overlap regime is less favorable for LLM inference (especially decode coupled with KV-cache offloading); we discuss this asymmetry in Appendix~\ref{app:llm}.} 

\subsection{Scaling over Frame Size and Batch Size}
\label{subsec:eval-scaling}

This evaluation has two goals: (i) evaluating how \name performs across a range of per-GPU compute workloads, and (ii) validating the overlap model developed in Section~\ref{subsec:design-overlap-model}. For the video models WanVideo and HunyuanVideo, we sweep the number of output frames, which changes the post-patchification sequence length $S$. For the image model Flux, we sweep the batch size $B$ instead, which changes the block-level compute workload $F_{\text{block}}$ through the linear-in-$B$
dependence in the FLOP model. Across all settings we report the average denoising step time and GPU peak memory.

\begin{figure}[t]
\centering
\includegraphics[width=\linewidth]{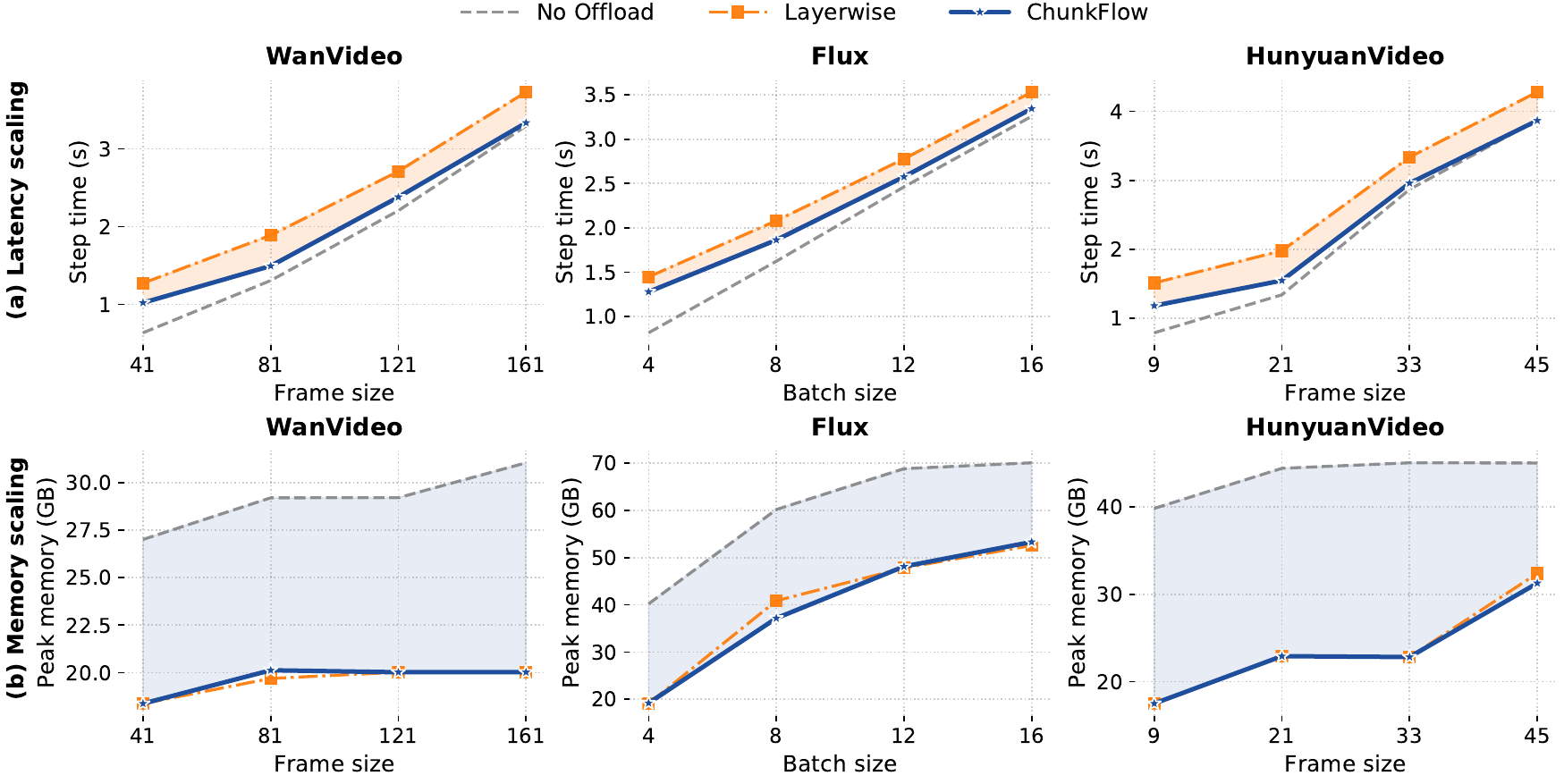}
\caption{Denoising step time (top) and GPU peak memory (bottom) across frame and batch sizes}
\label{fig:scaling}
\vspace{-10pt}
\end{figure}

\paragraph{Peak memory.}
Although \name operates at chunk rather than whole-layer granularity, the
bottom row of Figure~\ref{fig:scaling} shows that its peak memory is
essentially the same as Layerwise and remains
substantially below No Offload. At the third configuration point of
each model---past the predicted $F^\star$---\name reduces peak memory
by $30.5\%$, $30.1\%$, and $49.3\%$ over No Offload on
WanVideo, Flux, and HunyuanVideo, respectively, while keeping
denoising step time within a small margin of No Offload baseline. The finer granularity therefore introduces no memory overhead. In a few cases, such as Flux at batch size $8$, \name in fact consumes less
memory than layerwise offloading. This comes from allocating chunk buffers at
a fixed size, which stabilizes the allocator's request pattern and reduces
memory fragmentation compared to allocating whole-layer-sized blocks.

\paragraph{Step time.}
Figure~\ref{fig:scaling} (top row) shows that \name consistently
achieves lower denoising step time than Layerwise across all three
models and all configurations. In the figure, the orange shaded
regions in the top row mark the step-time speedup of \name over
Layerwise, and the blue shaded regions in the bottom row mark the
peak-memory savings of \name over No Offload. Concretely, \name
delivers up to $1.26\times$, $1.13\times$, and $1.28\times$ step-time
speedup over Layerwise on WanVideo, Flux, and HunyuanVideo,
respectively. As the frame size or batch size grows, \name's step
time steadily approaches No Offload, matching the prediction of
Section~\ref{subsec:design-overlap-model}: a larger $F_{\text{block}}$
extends the compute window available to hide the prefetch overhead.

We validate $F^\star$ derived in
Section~\ref{subsec:design-overlap-model}. The hardware constants
$P_{\text{peak}}$ and $BW_{\text{h2d}}$ are read from the platform
specification, and 
$\eta_{\text{comp}}$ and $\eta_{\text{pref}}$ are obtained from
offline profiling on the same platform; the exact values are listed in
Table~\ref{tab:calibration} of Appendix~\ref{app:calibration}.
Plugging these values into $F^\star$ yields thresholds that correspond
to approximately $121$ frames on WanVideo, batch size $12$ on Flux,
and $33$ frames on HunyuanVideo---that is, the third configuration
point in each panel. Starting from this point our step time is nearly
indistinguishable from No Offload, which is consistent with the
model. The measured gap continues to shrink slightly even beyond
$F^\star$, indicating that the first-order model is accurate to within
a small margin. The residual discrepancy is expected: the calibration
factors $\eta_{\text{comp}}$ and $\eta_{\text{pref}}$ are picked
empirically and not fitted per configuration, and \name itself
introduces a small amount of synchronization overhead (quantified in Section~\ref{subsec:eval-breakdown}) that is not explicitly modeled.

\subsection{Step Time Breakdown}
\label{subsec:eval-breakdown}

\begin{figure}[!t]
\centering
\begin{minipage}[t]{0.62\linewidth}
  \centering
  \includegraphics[width=\linewidth]{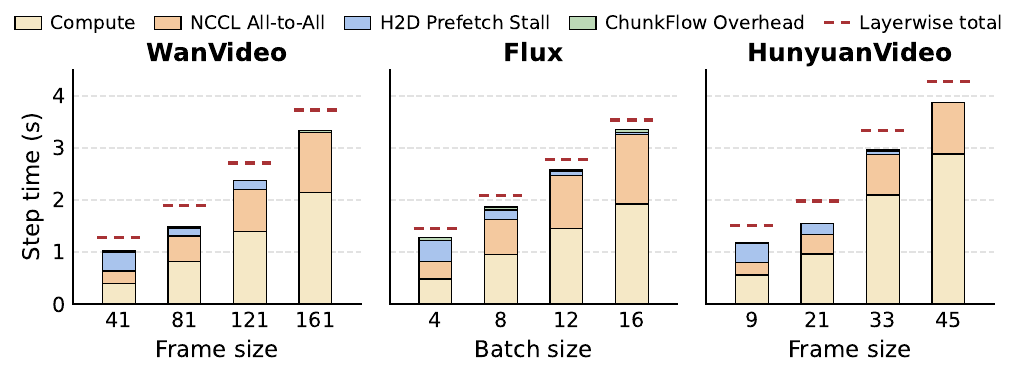}
  \caption{Per-step latency decomposition across frame and batch sizes with \name.}
  \label{fig:breakdown}
\end{minipage}\hfill
\begin{minipage}[t]{0.36\linewidth}
  \centering
  \includegraphics[width=\linewidth]{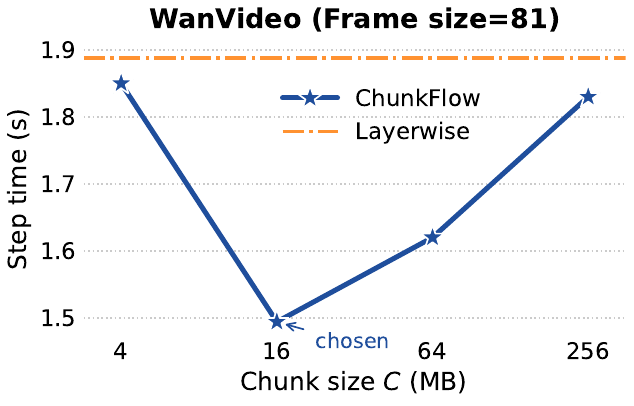}
  \caption{Step time vs.\ chunk size $C$ on WanVideo at $81$ frames.}
  \label{fig:chunksize}
\end{minipage}
\vspace{-15pt}
\end{figure}

To attribute \name's speedup over Layerwise to specific cost
components, we decompose \name's per-step latency at each swept
configuration into four parts: compute, NCCL all-to-all, H2D prefetch
stall, and \name runtime overhead (Figure~\ref{fig:breakdown}). Each
bar's height equals \name's step time. A dashed red marker above each
bar shows Layerwise's step time with the same configuration; the gap
between the bar top and marker is the PCIe contention stall
introduced in Section~\ref{subsec:background-communication-path} that
Layerwise pays and that \name eliminates. 

\paragraph{H2D prefetch stall shrinks with workload.}
Recapping the trend discussed in
Section~\ref{subsec:eval-scaling}: the H2D prefetch stall is large at
the smallest configuration of each model and shrinks as $n$ or
$b$ grows, becoming a thin sliver at the largest configuration. This
matches the prediction of the overlap model in
Section~\ref{subsec:design-overlap-model}---a wider compute window overlaps more of the prefetch. 

\paragraph{\name's runtime overhead is small.}
\name's own overhead in Figure~\ref{fig:breakdown} comes from two
sources. First, the runtime synchronizes the compute stream and the
prefetch worker---signaling a pause request at each communication
boundary and a resume request after the collective completes---which
adds a small fixed cost per collective invocation. Second, as
illustrated in Figure~\ref{fig:timeline}(c), when the compute stream
signals a pause, the copy stream finishes the in-flight chunk before
yielding, so the collective waits up to one chunk-transfer time. In
practice both contributions remain small: across all three models and
all swept configurations, \name's overhead stays within about $2\%$ of
the total step time, far smaller than the PCIe contention stall it
eliminates.

\subsection{Chunk Size Selection}
\label{subsec:eval-chunksize}

\name's chunk size $C$ controls the granularity at which the prefetch
worker yields to NCCL collectives, and the right value balances two
opposing effects. Figure~\ref{fig:chunksize} sweeps
$C \in \{4, 16, 64, 256\}$~MB on WanVideo at $81$ frames. At
$C=4$~MB, the per-chunk DMA startup cost is amortized over too few
bytes, effective PCIe bandwidth drops, and step time degrades close
to Layerwise. As $C$ grows to $64$~MB and then $256$~MB, the chunks become
progressively coarser and expose fewer yield points to collectives,
so the prefetch behaves more and more like a whole-layer transfer
and the schedule gradually falls back toward Layerwise. The
chunk-tail stall incurred at each yield---bounded by
$C / (\eta_{\text{pref}} BW_{\text{h2d}})$ as discussed in
Section~\ref{subsec:design-chunked-prefetch}---also grows linearly with $C$,
amplifying the fixed yielding cost on top of the lost overlap. $C =
16$~MB sits at the sweet spot between these two regimes and is the
operating point we use in all other experiments.

\subsection{Memory--Latency Tradeoff via Partial Residency}
\label{subsec:eval-resident}

We evaluate the partial parameter residency mechanism introduced
in Section~\ref{subsec:design-partial-residency}. For each model we deliberately
pick the smallest configuration from the previous experiment---$41$ frames on
WanVideo, batch size $4$ on Flux, and $9$ frames on
HunyuanVideo---because at this scale the compute window is short enough such that a
substantial portion of each layer's prefetch is exposed to the critical
path under full offloading (\textcolor{checked}{i.e., no residency}), leaving meaningful headroom for residency to 
recover. Starting from the fully-offloaded setting (\name in
Figure~\ref{fig:scaling}, labelled as $0\%$ residency here), we progressively
keep a $20\%$, $40\%$, and $60\%$ fraction of each layer's parameter chunks
resident in GPU memory and prefetch only the remaining chunks. Because \name's runtime manages parameters as fixed-size chunks, the residency ratio is enforced
at chunk granularity rather than at the coarser module or whole-layer
granularity, so the actual resident fraction closely tracks the target value.

\begin{figure}[!t]
\centering
\includegraphics[width=\linewidth]{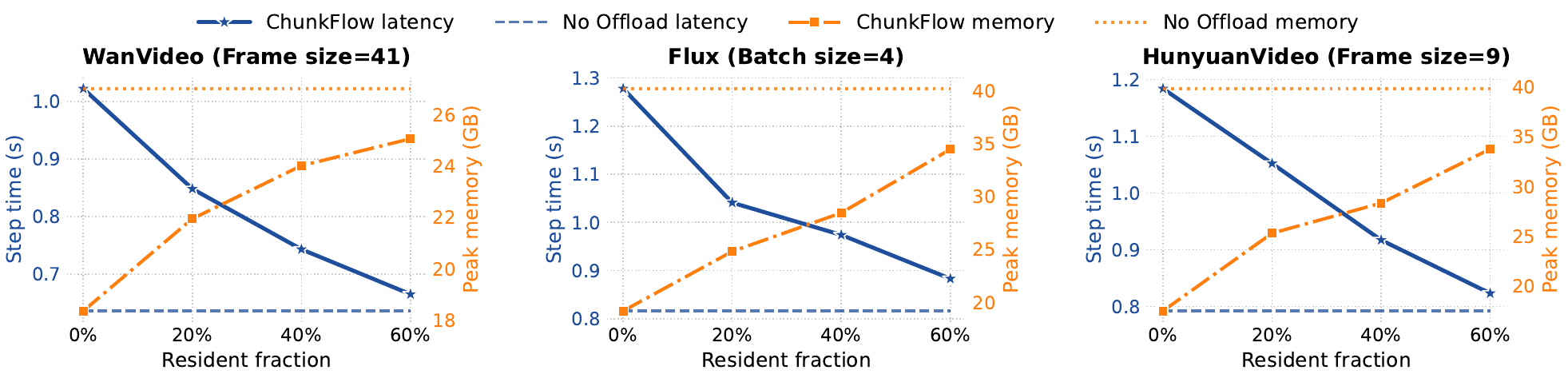}
\caption{Denoising step time and GPU peak memory at the smallest
configuration.}
\label{fig:resident}
\vspace{-20pt}
\end{figure}

Figure~\ref{fig:resident} illustrates the memory--latency tradeoff. 
As the resident fraction
grows, the prefetch volume per layer shrinks, and denoising step time
monotonically decreases toward No Offload. Correspondingly, peak
memory rises monotonically from the fully-offloaded level toward No Offload, since more parameters must sit in GPU memory simultaneously. The two trends are strictly opposed, giving a continuous spectrum between
minimum memory (full offload) and minimum latency (full residency). At only $60\%$ residency, the step time of all three models comes
within a very small margin of No Offload, indicating that the
prefetch of the remaining $40\%$ of parameters can be almost fully hidden by
the compute window. At this point \name still reduces peak memory by
$7.2\%$, $14.3\%$, and $15.2\%$ over No Offload on WanVideo,
Flux, and HunyuanVideo, respectively. In other words, even in the
low-compute regime where full offloading leaves visible overhead, \name
can recover near-zero step-time overhead while still delivering a meaningful
reduction in peak memory. In
practice, this allows the user to dial in any point along the curve to meet a
latency target under a given memory budget, or conversely to minimize memory
while staying within a target step time.

\section{Related Work}
\label{sec:related}

\textbf{Memory-efficient inference via offloading.} Existing efforts save GPU memory by offloading weights or activations to host or NVMe storage~\cite{hpca21:ren,atc21:zerooffload,lan2025zenflowenablingstallfreeoffloading,hpca24_dynn,sc24_cxl,kim2025costefficientllmservingcloud}.
ZeRO-Infinity~\cite{zeroinfinity2021} pioneered hierarchical CPU/NVMe
offloading. FlexGen~\cite{flexgen2023} formulates LLM inference as a tensor placement
problem and searches for high-throughput offloading schedules under 
GPU memory budgets. vLLM~\cite{kwon2023efficient} introduces paged KV-cache
management built on tensor parallelism~\cite{megatron2019}, and 
supports group-wise weight offloading at layer granularity.
MoE-Lightning~\cite{moelightning2024} targets MoE inference and develops a
CPU--GPU--I/O pipelining schedule with paged weights to overlap expert
transfer with computation. Production-quality inference systems such as
SGLang~\cite{sglang2024} and vLLM~\cite{kwon2023efficient} rely on naive whole-layer
prefetching. 
They do not model the contention when distributed collectives and H2D transfers share a PCIe fabric. 

\textbf{Efficient diffusion inference.} Diffusion model inference is dominated by repeated evaluations of a large
denoiser, and a long line of work attacks this bottleneck along two
complementary axes. The first axis reduces per-step compute or the
number of steps. Faster ODE/SDE samplers~\cite{ddim2021,dpmsolver2022,edm2022}
cut the number of denoising steps from hundreds to tens, and step-distillation
methods such as consistency models~\cite{consistency2023} push generation
toward a few or even a single step. Architectural efficiency is improved
through better DiT designs~\cite{pixart2024,mmdit2024} and post-training
optimizations including weight quantization~\cite{qdiffusion2023} and
feature caching that reuses high-level activations across adjacent denoising
steps~\cite{deepcache2024,learning2cache2024}. The second axis is
distributing computation across multiple GPUs.
DistriFusion~\cite{distrifusion2024} partitions the latent into spatial
patches and reuses stale feature maps to amortize cross-patch communication;
PipeFusion~\cite{pipefusion2024} extends this with a patch-level pipeline
parallel schedule; xDiT~\cite{xdit2024} integrates sequence parallelism,
patch-level pipeline parallelism, and CFG parallelism into a unified
inference engine; 
these two lines of work reduce or distribute computation, while we 
study how to combine layerwise weight offloading with distributed execution. 

\section{Conclusions}
\label{sec:conclusion}

We revisited layerwise offloading for DiT inference as a
prefetch--communication co-scheduling problem. Guided by a first-order
overlap model that defines a critical compute workload $F^\star$, we
designed a chunk-granular offloading runtime with two mechanisms:
communication-aware chunked prefetching, which \textcolor{checked}{eliminates PCIe contention by yielding} the PCIe Rx to
collective communication at chunk boundaries \textcolor{checked}{on PCIe-based nodes}, and chunk-granular partial
parameter residency, \textcolor{checked}{a general mechanism that} trades GPU memory for prefetch volume \textcolor{checked}{whenever the compute window is too short, regardless of interconnect}.
Experiments on WanVideo, Flux, and HunyuanVideo on two H100 GPUs
over PCIe with Ulysses sequence parallelism show that \name consistently
accelerates layerwise offloading, matches the no-offload baseline \textcolor{checked}{once the workload is large enough} 
while retaining substantial peak-memory savings, 
and recovers near-zero step-time overhead for small workloads through partial residency.


\bibliographystyle{plainnat}
\bibliography{han, li}

\clearpage
\appendix

\section{Roofline View of the Overlap Threshold}
\label{app:roofline}

The critical compute workload $F^\star$ derived in
Section~\ref{subsec:design-overlap-model} admits a natural geometric
interpretation. Define the per-layer \emph{operational intensity} as the
ratio of block-level FLOPs to the parameter bytes that must be prefetched
before the block executes,
\begin{equation}
I_{\text{block}}(B, S) \;:=\; \frac{F_{\text{block}}(B, S)}{B_{\text{pref}}}
\quad \text{(FLOPs/byte)}.
\end{equation}
The overlap condition $F_{\text{block}} \ge F^\star$ is then equivalent to
$I_{\text{block}} \ge I^\star$, where
\begin{equation}
I^\star
\;=\;
\frac{F^\star}{B_{\text{pref}}}
\;=\;
\frac{\eta_{\text{comp}} P_{\text{peak}}}{\eta_{\text{pref}} BW_{\text{h2d}}}.
\end{equation}
\begin{wrapfigure}{r}{0.32\linewidth}
\vspace{-0.6em}
\centering
\includegraphics[width=\linewidth]{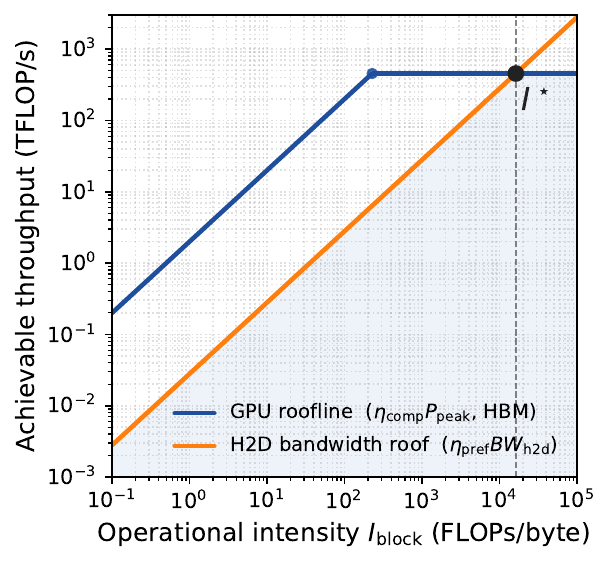}
\caption{HRM with the GPU roofline and the host-to-device bandwidth roof.}
\label{fig:roofline}
\vspace{-0.4em}
\end{wrapfigure}
Plotted as a roofline (Figure~\ref{fig:roofline}), $I^\star$ is the turning
point between the calibrated GPU compute roof
$\eta_{\text{comp}} P_{\text{peak}}$ and the host-to-device bandwidth roof
of slope $\eta_{\text{pref}} BW_{\text{h2d}}$. Workloads whose operational
intensity lies to the right of $I^\star$ admit full prefetch overlap;
workloads to its left are H2D-bound and expose a portion of the transfer on
the critical path.

\paragraph{Connection to the Hierarchical Roofline Model.}
The Hierarchical Roofline Model (HRM)~\cite{moelightning2024} provides a
general framework for analyzing computations that span multiple memory
hierarchies, and has been used to guide policy search for offloaded MoE
inference. Our analytical model can be viewed as an adaptation of HRM to
the specific question of prefetch--compute overlap in layerwise
offloading: by specializing the operational intensity to the per-block
prefetch volume $B_{\text{pref}}$ and folding empirical calibration
factors into the two roofs, the same geometry yields a direct view of
when the next layer's prefetch can be hidden behind the current layer's
compute.

\section{Per-Block Compute Model: DiT and MM-DiT}
\label{app:block-compute}

This appendix instantiates the overlap model of
Section~\ref{subsec:design-overlap-model} for the two block designs used
by our evaluated models: the standard DiT block (WanVideo) and the
MM-DiT block (Flux, HunyuanVideo). Both follow the same template---
block-level FLOPs summed over the dominant operators, an empirical
calibration factor $\eta_{\text{comp}}$, and the same critical compute
workload threshold $F^\star$---but differ in the per-term FLOPs because
MM-DiT processes the image and text token streams jointly in a single
attention call rather than via a self-attention and a separate text
cross-attention.

We reuse the notation of Section~\ref{subsec:design-overlap-model}: $B$
is the batch size, $S$ the image/video token length after
patchification, $L_{\text{ctx}}$ the text context length, $d$ the
hidden dimension, and $f$ the FFN expansion. FLOP counts below are written in global (across-GPU) form; under distributed execution each GPU performs only a fraction of them, and the per-GPU breakdown depends on the parallelism scheme. Tensor parallelism with degree $T_{\text{TP}}$ and Ulysses sequence parallelism with degree $T_{\text{SP}}$ both partition the head dimension, so every FLOP term below is divided by the corresponding degree on a per-GPU basis. Token-partitioned sequence parallelism (e.g., ring attention) instead splits the sequence $S$ across devices, so the projection and MLP terms still divide by $T_{\text{SP}}$ while the attention term is divided according to the variant's load-balancing strategy. In all cases the structure of the overlap model is unchanged---only the per-GPU $F_{\text{block}}$ and hence $F^\star$ are reparameterized---and the chunked prefetching and partial residency mechanisms apply unmodified. Residual losses (e.g., load imbalance, kernel-launch overhead) are absorbed into the calibration factor $\eta_{\text{comp}}$.

\paragraph{DiT block.}
A standard DiT block serially interleaves self-attention,
cross-attention to the text context, and an MLP. The dominant FLOPs
are
\begin{align}
F_{\text{self-proj}}  &= 8 B S d^2,                              \\
F_{\text{self-attn}}  &= 4 B S^2 d,                              \\
F_{\text{cross-proj}} &= 4 B S d^2 + 4 B L_{\text{ctx}} d^2,     \\
F_{\text{cross-attn}} &= 4 B S L_{\text{ctx}} d,                 \\
F_{\text{mlp}}        &= 4 B S d f,
\end{align}
where ``self-proj'' covers the QKV and output projection of the
self-attention, and ``cross-proj'' covers the video-side query, the
text-side key/value, and the output projection of the cross-attention.
Summing these dominant terms gives the per-block FLOP count
\begin{equation}
\label{eq:flop-dit}
F_{\text{block}}^{\text{DiT}}(B, S)
=
F_{\text{self-proj}}
+ F_{\text{self-attn}}
+ F_{\text{cross-proj}}
+ F_{\text{cross-attn}}
+ F_{\text{mlp}}.
\end{equation}

The MM-DiT denoiser is built from two block types (double-stream and
single-stream) that we model separately and then combine via an
averaged block size.

\paragraph{Double-stream block.}
A double-stream block (e.g., \texttt{transformer\_blocks} in Flux,
\texttt{double\_blocks} in HunyuanVideo) maintains separate image-side and
text-side weights for QKV projection, output projection, and the FFN, but
the two streams are concatenated along the token dimension before the joint
attention and split afterwards. The dominant FLOPs are
\begin{align}
F_{\text{img-proj}} &= 8 B S d^2,                              \\
F_{\text{txt-proj}} &= 8 B L_{\text{ctx}} d^2,                 \\
F_{\text{joint-attn}} &= 4 B (S + L_{\text{ctx}})^2 d,         \\
F_{\text{img-mlp}}    &= 4 B S d f,                            \\
F_{\text{txt-mlp}}    &= 4 B L_{\text{ctx}} d f,
\end{align}
where the ``proj'' terms cover both QKV and output projection on each side
($4 d^2$ each, hence the leading coefficient $8$). Summing,
\begin{equation}
F_{\text{dbl}}(B, S) = F_{\text{img-proj}} + F_{\text{txt-proj}}
  + F_{\text{joint-attn}} + F_{\text{img-mlp}} + F_{\text{txt-mlp}}.
\end{equation}
Compared with the DiT block, the quadratic-in-$S$ contribution is replaced
by a quadratic-in-$(S+L_{\text{ctx}})$ contribution from joint attention,
and the cross-attention term disappears.

\paragraph{Single-stream block.}
A single-stream block (e.g., \texttt{single\_transformer\_blocks} in Flux,
\texttt{single\_blocks} in HunyuanVideo) concatenates the image and text
tokens at block entry and processes them as one stream. It uses a single
fused linear $\mathtt{linear}_1$ that produces both QKV and the MLP
up-projection, and a single fused linear $\mathtt{linear}_2$ that absorbs
the attention output and the MLP down-projection:
\begin{align}
F_{\text{sng-lin1}} &= 2 B (S + L_{\text{ctx}}) d (3 d + f), \\
F_{\text{sng-attn}} &= 4 B (S + L_{\text{ctx}})^2 d,         \\
F_{\text{sng-lin2}} &= 2 B (S + L_{\text{ctx}}) (d + f) d,
\end{align}
\begin{equation}
F_{\text{sng}}(B, S) = F_{\text{sng-lin1}} + F_{\text{sng-attn}} + F_{\text{sng-lin2}}.
\end{equation}

\paragraph{Block-averaged compute time.}
Let $N_{\text{d}}$ and $N_{\text{s}}$ be the number of double-stream and
single-stream blocks in the model. We collapse the two block types into a
single average block when applying the overlap condition:
\begin{equation}
\bar{F}_{\text{block}}(B, S)
= \frac{N_{\text{d}} F_{\text{dbl}}(B, S) + N_{\text{s}} F_{\text{sng}}(B, S)}
       {N_{\text{d}} + N_{\text{s}}},
\qquad
\bar{T}_{\text{comp}}(B, S)
= \frac{\bar{F}_{\text{block}}(B, S)}{\eta_{\text{comp}} P_{\text{peak}}}.
\end{equation}

\paragraph{Prefetch model.}
The prefetch model carries over without change. The per-block parameter
volume $B_{\text{pref}}$ used in $T_{\text{pref}} = B_{\text{pref}} /
(\eta_{\text{pref}} BW_{\text{h2d}})$ is now an average across the two
block types: a double-stream block contributes roughly $\beta(20 d^2 + 4 d
f)$ bytes (two attention projection sets, two MLPs, plus modulation), and a
single-stream block contributes roughly $\beta(7 d^2 + 2 d f)$ bytes (the
fused $\mathtt{linear}_1$ and $\mathtt{linear}_2$ together with one
modulation), where $\beta$ is the parameter byte width.

\paragraph{Critical compute workload.}
With $\bar{F}_{\text{block}}$ and the average $B_{\text{pref}}$ above, the
overlap condition $\bar{F}_{\text{block}}(B, S) \ge F^\star$ and the
critical compute workload $F^\star$ defined in
Section~\ref{subsec:design-overlap-model} apply unchanged. Increasing $B$
or $S$ continues to grow $\bar{F}_{\text{block}}$ monotonically, with the
same regime-boundary interpretation as the DiT case.

\paragraph{Timing under distributed offloading.}
Both Flux and HunyuanVideo place the sequence-parallel collectives inside
the joint attention call of each block (as well as within the joint
attention of the single-stream block, which operates on the concatenated
image+text sequence). The block-level timing therefore mirrors
Figure~\ref{fig:timeline}: in the layerwise-offload schedule, the
in-flight H2D prefetch can block the in-block all-to-all at the GPU Rx and
expose prefetch latency on the critical path; in \name's chunked-prefetch
schedule, the prefetch is paused at chunk boundaries and resumes after the
collective completes. Because the chunked-prefetch and partial-residency
mechanisms of Sections~\ref{subsec:design-chunked-prefetch}
and~\ref{subsec:design-partial-residency} operate on parameter chunks
without reference to the block's internal structure, they apply to MM-DiT
blocks with no structural changes.

\section{Numerical Instantiation and $F^\star$ Predictions}
\label{app:calibration}

This appendix instantiates the overlap model of Section~\ref{subsec:design-overlap-model} on our evaluation platform: it lists the hardware constants, calibration factors, and per-model parameters, then plugs them into the overlap condition to derive the predicted critical configuration for each model.

Our evaluation node has two H100 PCIe GPUs that share the host PCIe root
complex, running BF16 inference with Ulysses sequence parallelism degree
$2$ and chunked prefetching. The hardware constants $P_{\text{peak}}$ and
$BW_{\text{h2d}}$ are fixed by the platform and are read directly from
the device specification: $P_{\text{peak}}$ is the peak BF16 Tensor Core
throughput of a single H100 PCIe in dense (no-sparsity) mode, and
$BW_{\text{h2d}}$ is the PCIe Gen5 $\times 16$ effective bandwidth
($\approx 63$ GB/s) halved because the two GPUs share the host PCIe
root.

The calibration factors $\eta_{\text{comp}}$ and $\eta_{\text{pref}}$ are
obtained from offline profiling on the same platform:
$\eta_{\text{comp}}$ from the BF16 FLOP/s achieved on a single DiT
block on each of the three evaluated models, and
$\eta_{\text{pref}}$ from the H2D throughput measured under SGLang's
layerwise prefetch. The per-model values of $\eta_{\text{comp}}$ differ
slightly across the three models; to keep the overlap model simple, we
use the average. The four hardware and calibration constants are summarized in
Table~\ref{tab:calibration}.

\begin{table}[!ht]
\centering
\caption{Hardware constants and empirical calibration factors used to
evaluate $F^\star$ in Section~\ref{subsec:eval-scaling}.}
\label{tab:calibration}
\begin{tabular}{l c}
\toprule
Symbol & Value \\
\midrule
$P_{\text{peak}}$       & $756$ TFLOP/s \\
$BW_{\text{h2d}}$       & $31.5$ GB/s   \\
$\eta_{\text{comp}}$    & $0.60$        \\
$\eta_{\text{pref}}$    & $0.89$        \\
\bottomrule
\end{tabular}
\end{table}

The FLOP counts $F_{\text{block}}$ in
Section~\ref{subsec:design-overlap-model} and
Appendix~\ref{app:block-compute} are the global per-block FLOPs,
i.e., summed over the entire batch and sequence before sequence-parallel
partitioning; under SP degree $2$ each GPU performs $F_{\text{block}}/2$
FLOPs. WanVideo uses the standard DiT block of
Section~\ref{subsec:design-overlap-model} with $30$ blocks, while Flux
and HunyuanVideo use the MM-DiT block of the same appendix
with $(N_d, N_s) = (19, 38)$ and $(20, 40)$ double-/single-stream blocks
respectively. The other formula parameters and the effective per-block
prefetch volume $B_{\text{pref}}$ are summarized in
Table~\ref{tab:per-model-params}: $B$ is the batch size (swept for Flux;
fixed at $1$ for the two video models), $d$ the hidden dimension, $f$
the FFN intermediate dimension, $L_{\text{ctx}}$ the text context
length, and $S$ the image-token sequence length. The $S$ formula in
each row reflects the model's input resolution---WanVideo at $704
\times 1280$, Flux at $1024 \times 1024$, and HunyuanVideo at $720
\times 1280$---together with its VAE compression and patchification.
$B_{\text{pref}}$ is obtained from offline profiling on each model.

\begin{table}[!ht]
\centering
\caption{Per-model formula parameters and effective per-block prefetch
volume $B_{\text{pref}}$.}
\label{tab:per-model-params}
\begin{tabular}{l c c c c l c}
\toprule
Model & $B$ & $d$ & $f$ & $L_{\text{ctx}}$ & Sequence length $S$ & $B_{\text{pref}}$ \\
\midrule
WanVideo      & $1$ & $3072$ & $14336$ & $512$ & $220(n{+}3)$ & $520$\,MB \\
Flux          & $b$ & $3072$ & $12288$ & $512$ & $4096$       & $465$\,MB \\
HunyuanVideo  & $1$ & $3072$ & $12288$ & $161$ & $900(n{+}3)$ & $675$\,MB \\
\bottomrule
\end{tabular}
\end{table}

Plugged into the per-GPU overlap condition $T_{\text{comp}} =
T_{\text{pref}}$ together with the constants in
Table~\ref{tab:calibration}, the values in
Table~\ref{tab:per-model-params} yield predicted critical configurations
of $n^\star \approx 119.2$ frames for WanVideo, $b^\star \approx 11.5$
for Flux, and $n^\star \approx 34.5$ frames for HunyuanVideo. Rounding
each to the nearest valid swept configuration in
Section~\ref{subsec:eval-scaling} gives $121$ frames, batch $12$, and
$33$ frames respectively, which are the third configuration we chose for
each model.

\section{Applicability to LLM Inference}
\label{app:llm}

The design of \name---the overlap model
(Section~\ref{subsec:design-overlap-model}), communication-aware
chunked prefetching (Section~\ref{subsec:design-chunked-prefetch}),
and partial parameter residency
(Section~\ref{subsec:design-partial-residency})---is not
architecturally tied to diffusion transformers: each component depends
only on per-GPU compute throughput, host-to-device bandwidth, and the
layered structure of the forward pass, all of which are shared by any
transformer-based inference workload. We focus on DiT inference rather
than LLM inference because the underlying overlap regime is far more
favorable for DiT, as outlined below.

\paragraph{Long per-step compute window favors DiT.}
DiT inference processes the entire post-patchification token sequence in
every denoising step; the per-block compute cost $F_{\text{block}}$
scales with the global sequence length $S$
(Section~\ref{subsec:design-overlap-model}). For high-resolution image
and video models, $S$ routinely reaches $10^5$--$10^6$ tokens, pushing
$F_{\text{block}}$ above the critical compute workload $F^\star$ in
the configurations we evaluate. LLM inference operates at a much
smaller scale: prefill sequences are typically only on the order of
$10^3$--$10^4$ tokens (e.g., the default per-batch token budget of
vLLM~\cite{kwon2023efficient} is $2048$), and decode produces a single
new token per request per step, so the effective per-step sequence
length equals the batch size, which is generally smaller than the
prefill length. The corresponding $F_{\text{block}}$ is therefore
substantially smaller than DiT's, placing typical LLM workloads well
below $F^\star$, with decode being the most extreme case.

\paragraph{Memory-bound regimes (decode, MoE) make prefetch even harder.}
Two important LLM regimes are dominated by memory traffic rather than
tensor-core arithmetic: dense-model decode and MoE inference, where
the per-layer compute is bottlenecked by HBM weight (and, in decode,
KV-cache) reads rather than peak FLOP/s. Since HBM bandwidth exceeds
PCIe bandwidth by roughly an order of magnitude (e.g., $\sim
3$\,TB/s vs $64$\,GB/s on H100 PCIe), an H2D prefetch of the same
weights over PCIe is substantially slower than the HBM-bound compute
it must hide behind, so $T_{\text{pref}} \gg T_{\text{comp}}$ once
layerwise offloading is enabled. Recovering overlap via partial
residency (Section~\ref{subsec:design-partial-residency}) would
require a resident fraction so high that it approaches the no-offload
baseline, eroding the memory-saving rationale for offloading in the
first place.

\paragraph{KV-cache offloading further compresses the prefetch bandwidth budget.}
In memory-constrained LLM deployments, KV-cache offloading is commonly
enabled alongside weight offloading: the KV cache is staged in host
memory and streamed to the GPU as needed. This adds a second stream
of H2D traffic in both prefill and decode, competing with
model-weight prefetch on the same PCIe path and effectively shrinking
the prefetch calibration factor $\eta_{\text{pref}}$ in the overlap
model. The result is to push $F^\star$ even higher, exacerbating the
compute-bound regime described above.

\paragraph{Summary.}
All three components of \name---the overlap model, chunk-granular
pause/resume, and partial residency---are general and could be applied
to LLM inference with no architectural change. Their practical
effectiveness, however, is governed by the overlap model itself: DiT
inference sits comfortably in the regime where prefetch can be hidden,
whereas LLM inference (decode in particular, especially with
concurrent KV-cache offloading) sits deep in the compute-bound regime
where neither chunked prefetching nor moderate residency is
sufficient. A
satisfactory treatment of LLM inference under offloading would likely
require co-designing model-weight and KV-cache offloading scheduling,
which is beyond the scope of this work.

\section{Broader Impact}
\label{app:broader-impact}

The system optimizations proposed in this work reduce the GPU memory and
end-to-end latency of distributed diffusion transformer inference. The
direct positive consequences are lower hardware requirements and lower
energy cost for deploying large diffusion transformers, which can
democratize access to high-quality image and video generation and reduce
the carbon footprint of generative inference. We do not propose new
generative models or new training data, and our method is purely a
runtime-level optimization that leaves the underlying model and sampler
unchanged; consequently, it does not introduce new capabilities for
content generation. As with any infrastructure that lowers the cost of
diffusion model deployment, our work indirectly makes the existing
limitations of large generative models---including potential misuse for
disinformation or deepfakes, and existing biases inherited from training
data---easier to encounter at scale. Our method does not address these
issues, which we view as the responsibility of model developers, model
licensors, and downstream system operators rather than of the inference
runtime itself.


\end{document}